# New Characterization of Disorder Taming Spatiotemporal Chaos

*Feng Qi, Zhonghuai Hou, Houwen Xin**

*Department of Chemical Physics, University of Science and Technology of China, Hefei, Anhui, 230026, P. R. China*
.

**Abstract:** In the present letter, we introduce a new method to quantify the effect of disorder on spatiotemporal chaos [Y. Braiman, etc. Nature, 378, p465 (1995)]. Base on the autocorrelation function, we define a parameter to measure the effect of disorder. The results are intriguing and similar to the results obtained by J. Lindner etc. [J. Lindner, etc. Physics Letters A, 231, p164 (1997)]. Using this order parameter, the effect of disorder on spatiotemporal chaos in different conditions has also been discussed.

*PACS:* 05.45. + b; 64.60.Cn

*Author to whom correspondence should be addressed. Fax: 86-551-3606046, email: fengqi@mail.ustc.edu.cn

Recently, order induced by noise and disorder in nonlinear systems has become a topic of great interest, due to its theoretical significance and practical applications. Intriguing examples include stochastic resonance [1], taming spatiotemporal chaos by disorder [2], disorder-enhanced synchronization [3-4], etc.

Because the real systems always have some natural or ineluctable spread in features, so it is very important to study the disorder coupling systems. After demonstrating the effect of disorder taming spatiotemporal chaos [2], J. Lindner etc. introduced a way to quantify the effect of disorder on spatiotemporal chaos based on the system's largest Lyapunov exponent [5]. In this letter, we will introduce a new method, and get similar results.

Here we use the same simple example as in Ref 2. Considering a array of forced, damped, pendulum governed by the equation

$$ml_n^2 \ddot{\theta}_n + \gamma \dot{\theta}_n = -mgl_n \sin\theta_n + \tau' + \tau \sin\omega t + \kappa(\theta_{n+1} + \theta_{n-1} - 2\theta_n) \quad (1)$$

Where $n = 1, 2, \ldots N$ (in our numerically investigating, we let $N = 128$), and the boundary condition is free ($\theta_0 = \theta_1$ and $\theta_N = \theta_{N+1}$). For simplicity, the parameters used are the gravitational acceleration $g = 1.0$, mass of the pendulum $m = 1.0$, length $l_n = 1.0$, d.c. torque $\tau' = 0.7155$, a.c. torque $\tau = 0.4$, the angular frequency $\omega = 0.25$, and the damping $\gamma = 0.75$, $\kappa$ is the coupling between the two oscillators. We numerically integrate the equation (1) using a fourth-order Runge-Kutta technique with a time step $dt = 0.001$.

One may find that the isolated pendulum is chaotic for the default length $l = 1.0$, which is characterized by a positive Lyapunov exponent. For $l > 1.0$, the pendulum executes a libration in which it oscillates about its equilibrium position without overturning, i.e. the angel $\theta$ never exceeds $2\pi$. On the other hand for $l < 1.0$, the pendulum executes a whirling where the combined torques rotates the pendulum over the top and the angel $\theta$ past $2\pi$. An isolated pendulum of default length $l = 1.0$ displays a chaotic sequence of whirling and libration.

To compare with the results obtained by J. Lindner[5], we also disorder the system in three different types: random disorder, alternate binary disorder, and linear disorder. The random disorder means that the length of pendulums is a uniform distribution $[1-\sigma, 1+\sigma]$. The alternate binary disorder means that the pendulum lengths alternate between a short $1-\sigma$ and a long $1+\sigma$. The linear disorder means the lengths increase linearly form $1-\sigma$ to $1+\sigma$. In all case, the average length of the pendulum is equal to one, in the chaotic region.

Figure 1 describes the continuous spatiotemporal evolution of an array of 128 pendulums. Time passes from the left to right. The colors code the angular velocities of each pendulum. We omit the transient at the beginning of the evolution. Figure 1a is an example of the evolution of a regular array without disorder ($\sigma = 0$). Spatiotemporal chaos can be observed. If we apply one type of disorder on the array, we can see interesting results, which are depicted in other panels of figure 1 (Here, we use random disorder as an example). Figure 1b shows one example after introducing disorder of $\pm 30\%$ that means the pendulums lengths are uniformly distributed in the interval $[0.7, 1.3]$. There appears a relatively simple but nontrivial pattern after a short transient (not shown). It is periodic in space and time. If increasing the degree of disorder more, some segments are still periodic, and some segments lose regularity. It is shown in Fig 1c, in which $\pm 60\%$ disorder is introduced.

To characterize the behavior induced by disorder, we introduce a new quantity. It is based on the normalized autocorrelation function $c_i(\tau)$, defined by $c_i(\tau) = \dfrac{\left\langle \tilde{\dot{\theta}}_i(t)\tilde{\dot{\theta}}_i(t+\tau)\right\rangle}{\left\langle \tilde{\dot{\theta}}_i^{\,2}\right\rangle}$, where $\dot{\theta}_i(t)$ is the angular velocity of the ith pendulum, $\tau$ is the time delay, and $\tilde{\dot{\theta}}_i(t) = \dot{\theta}_i(t) - \left\langle \dot{\theta}_i\right\rangle$. The characteristic correlation time is then evaluated as $\tau_{i,c} = \dfrac{1}{T}\int_T c_i^{\,2}(\tau)d\tau$, following Pikovsky et al [6]. In the present case of limited and discrete sampling, $\tau_c$ is evaluated by $\tau_{i,c} = \frac{1}{N\Delta t}\sum_{k=1}^{N} c_i^2(\tau_k)\Delta t$, where $\tau_c = k\Delta t$ with $\Delta t$ being the sampling time, and $N$ is the longest delay-time increment $\Delta t$.

Then we defined an "order parameter" described by $\tau(\sigma) = [\langle \tau_{i,c} \rangle]$. Here $\langle \bullet \rangle$ denotes averaging over all the pendulums, $[\bullet]$ denotes averages over 100 different disorder realizations with the same $\sigma$. The more orderly pendulum oscillating is, the more pronounced its correlations time is. So the corresponding characteristic correlation time is large. One can see from figure 2 that the correlations are indeed much more pronounced for moderate fraction of random connections.

Figure 3a shows the dependence of $\tau(\sigma)$ on the disorder amplitude $\sigma$ when the random disorder is introduced. The $\tau(\sigma)$-$\sigma$ curve clearly shows the regularity maximum at $\sigma_{opt}$. Here the $\sigma_{opt} \approx 0.3$. This result is similar with that in Ref. 5. Figure 3b and 3c are $\tau(\sigma)$-$\sigma$ curves corresponding to the other two types of disorder. When the alternate binary disorder is applied, there also exists an optimal disorder amplitude, now the $\sigma_{opt} \approx 0.5$. When the linear disorder is applied, the $\tau(\sigma)$-$\sigma$ curves have multi-peaks. We also check the effect of disorder when the system has different coupling constant. In all cases, changing the coupling constant will not change the trend of the $\tau(\sigma)$-$\sigma$ curves. However, changing the coupling constant has different effects in different disorder cases. In the random disorder case, when using larger coupling constant, the optimal disorder amplitude is shifted to the higher value and the corresponding $\tau(\sigma)$ also has higher value, as shown in figure 3a. In the alternate binary disorder case, increasing the coupling constant hardly change the optimal disorder amplitude, larger coupling constant makes the $\tau(\sigma)$-$\sigma$ curve more peaked, as shown in figure 3b, and do not change the maximum of the curve. In the linear disorder case, increasing coupling constant just makes the $\tau(\sigma)$ slightly larger at the peaks of the curve, as shown in figure 3c.

To check the noise effect on the phenomenon of disorder inducing regular, we rewrite the equation (1)

in following form

$$ml_n^2\ddot{\theta}_n + \gamma\dot{\theta}_n = -mgl_n \sin\theta_n + \tau' + \tau\sin\omega t + \kappa(\theta_{n+1} + \theta_{n-1} - 2\theta_n) + \beta_n\Gamma_n(t) \quad (2)$$

where $\Gamma(t)$ is the noise term chosen from some distribution. We choose $\Gamma(t)$ as a Gaussian distribution, such that $\langle\Gamma_n(t)\rangle = 0$, $\langle\Gamma_m(t)\Gamma_n(t')\rangle = \delta_{mn}\delta(t-t')$. $\beta_n$ is noise intensity, here we let $\beta_n = \beta$. The results are shown in figure 4. One can see clearly that with increment of noise intensity, the order parameter $\tau(\sigma)$ will decrease. However, even with large noise intensity, there are still a maximum on the $\tau(\sigma)$-$\sigma$ curve. That means the effect of disorder on the coupled pendulums is robust.

The chaotic region around $l \approx 1$ has been carefully examined. It has an extent of $0.998 < l < 1.002$ [7], and it is indeed quite a narrow region. If the lengths of the pendulums are disordered, shorter or longer than average, some of the oscillators will be renovated from their chaotic band, separately undergo periodic motion, thus form periodic islands. The remaining chaotic clusters are forced a locking to the external drive by these periodic islands. It creates periodic solutions to the equations of motions. It is one of the possible mechanisms by which disorder may stabilize a chaotic array. So applying different types of disorder, different distributions of disorder, should have different results. One can see it from figure 3. The existence of an optimal disorder which maximum of the "order parameter" of the array strengthens again the qualitative analogy to the phenomenon of SR.

In this letter, we use a new measurement to characterize the effect of disorder on coupled system. Similar with results in Ref. 5, we also observed that the existence of an optimal disorder which let the system spatiotemporal evolution most regular. We compare results under different conditions, such as using different disorder types, changing coupling constant. We know that taming spatiotemporal chaos should still be useful in mode-locking applications, such as super-conducting Josephson arrays, semiconductor laser arrays etc., where any type of regular behavior is preferred to chaos. What is corresponding mechanism to different types of disorder? What will happen when using different types of coupling? What are the effects of disorder on the different complex coupling systems, called complex networks [8]? How can we use this advantage to technological application? Further theoretical and experimental work will be of great help to answer these questions.

**Acknowledgement:** This work is supported by the National Science Foundation of China (Grant No. 20173052). The authors like to thank J. F. Lindner for useful discussions and the copy of the algorit

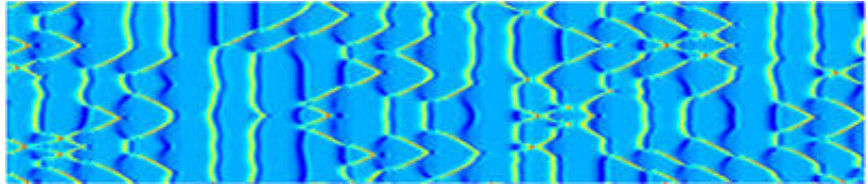

a)

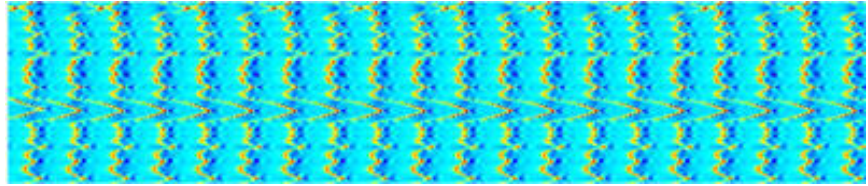

b)

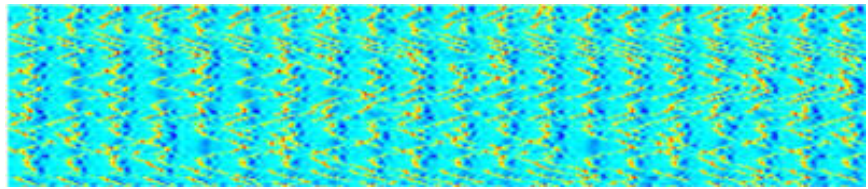

c)

**Fig. 1**. Spatiotemporal evolution of an array of 128 pendulums corresponding to different disorder amplitude $\sigma$. Time passes from the left to right. The colors code the angular velocities of each pendulum. a) $\sigma = 0.0$, b) $\sigma = 0.3$, c) $\sigma = 0.6$.

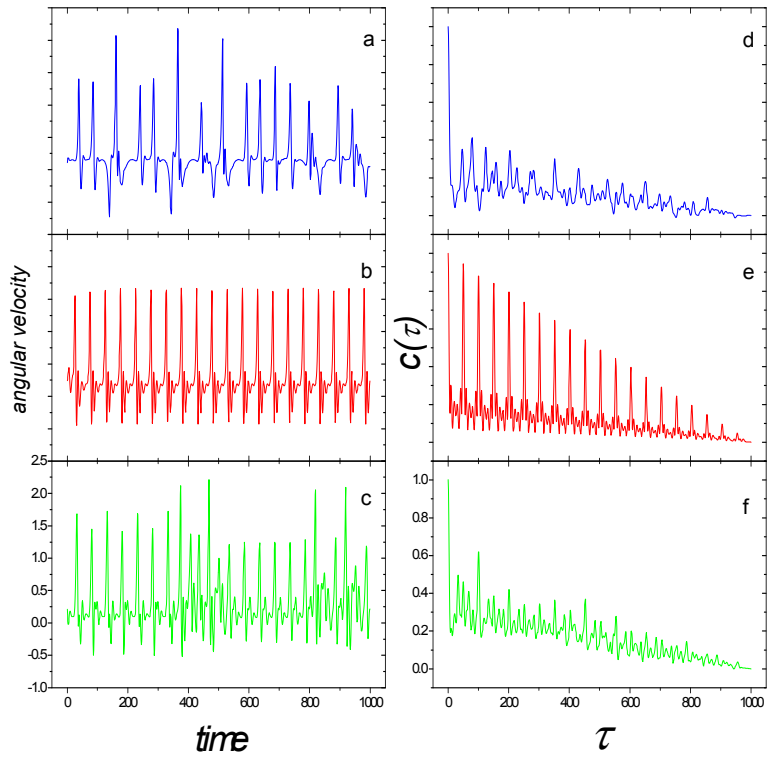

**Fig. 2.** Time series of the sixtieth pendulum (a, b, c) and the corresponding autocorrelation function time $c(\tau)$ (d, e, f) with different disorder amplitude. (a), (d) $\sigma = 0.0$; (b), (e) $\sigma = 0.3$; (c), (f) $\sigma = 0.6$.

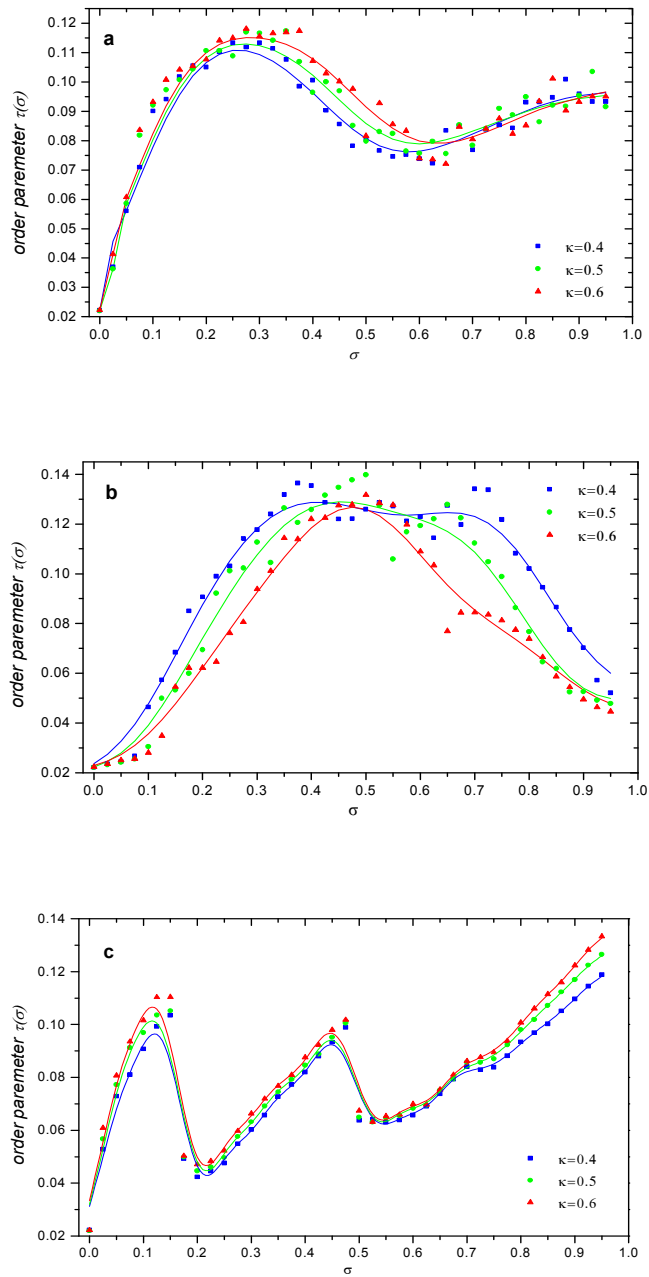

**Fig. 3.** The dependence of $\tau(\sigma)$ on the disorder amplitude $\sigma$ when introducing (a) random disorder, (b) alternate binary disorder, (c) linear disorder with coupling constant $\kappa$ of 0.4 (square), 0.5 (circle), 0.6 (up triangle). Solid lines are drawn as a guide the eye.

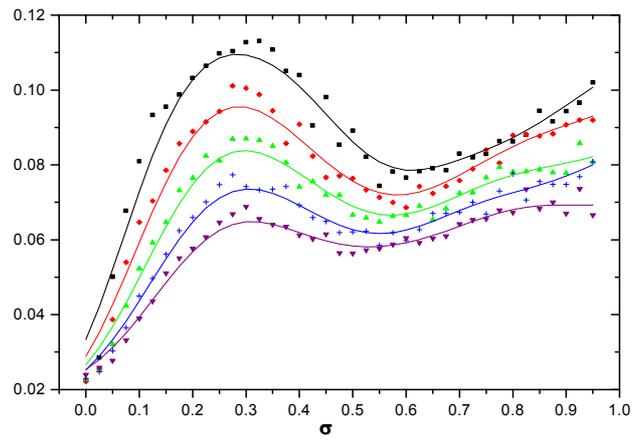

**Fig. 4.** The dependence of $\tau(\sigma)$ on the disorder amplitude $\sigma$ with noise intensity $\beta$ of 0.1 (square), 0.3 (diamond), 0.5 (up triangle), 0.7(cross), 0.9(down triangle). Solid lines are drawn as a guide the eye.